\documentclass[aps,prb,twocolumn,showpacs]{revtex4}
\usepackage{graphicx}

\begin{document}

\title{SrFeAsF as a parent compound for iron pnictide superconductors}

\author{Fei Han, Xiyu Zhu, Gang Mu, Peng Cheng,  and Hai-Hu Wen}\email{hhwen@aphy.iphy.ac.cn }

\affiliation{National Laboratory for Superconductivity, Institute of
Physics and Beijing National Laboratory for Condensed Matter
Physics, Chinese Academy of Sciences, P. O. Box 603, Beijing 100190,
China}

\begin{abstract}
We have successfully synthesized the fluo-arsenide SrFeAsF, a new
parent phase with the ZrCuSiAs structure. The temperature dependence
of resistivity and dc magnetization both reveal an anomaly at about
$T_{an}$ = 173 K, which may correspond to the structural and/or
Spin-Density-Wave (SDW) transition. Strong Hall effect and moderate
magnetoresistance were observed below $T_{an}$. Interestingly, the
Hall coefficient $R_H$ is positive below $T_{an}$, which is opposite
to the cases in the two parent phases of FeAs-based systems known so
far, i.e., LnFeAsO (Ln = rare earth elements) and (Ba,
Sr)Fe$_2$As$_2$ where the Hall coefficient $R_H$ is negative. This
strongly suggests that the gapping of the Fermi surface induced by
the SDW order leaves one of the hole pockets fully or partially
ungapped in SrFeAsF. Our data show that it is possible for the
parent phases of the arsenide superconductors to display dominant
carriers that are either electronlike or holelike.
\end{abstract} 
\maketitle

The discovery of superconductivity in the quaternary compound
LaFeAsO$_{1-x}$F$_x$ which is abbreviated as the FeAs-1111 phase,
has attracted great attentions in the fields of condensed matter
physics and material sciences.\cite{Kamihara2008} The family of the
FeAs-based superconductors has been extended rapidly. As for the
FeAs-1111 phase, most of the discovered superconductors are
characterized as electron-doped ones and the superconducting
transition temperature has been quickly raised to $T_c$ = 55$\sim$
56 K via replacing lanthanum with other rare earth
elements.\cite{XHCh,NLW,Pr52K,RenZA55K,CP,WangC} Meanwhile, the
first hole-doped superconductor La$_{1-x}$Sr$_{x}$FeAsO with $T_c
\approx$ 25 K was discovered,\cite{WenEPL,LaSr2} followed with the
observation of superconductivity in hole-doped
Nd$_{1-x}$Sr$_{x}$FeAsO\cite{NdSr} and
Pr$_{1-x}$Sr$_{x}$FeAsO.\cite{MuPrSr} Later on, (Ba,
Sr)$_{1-x}$K$_x$Fe$_2$As$_2$ which is denoted as FeAs-122 for
simplicity\cite{BaKparent,Rotter,CWCh}, and Li$_x$FeAs as an
infinite layered structure (denoted as FeAs-111) were
discovered.\cite{LiFeAs,LiFeAsChu,LiFeAsUK} It is assumed that the
superconductivity both in the FeAs-1111 phase and FeAs-122 phase is
intimately connected with a Spin-Density-Wave (SDW) anomaly in the
FeAs layers.\cite{SDW,BaKparent} For undoped LaFeAsO, an SDW-driven
structural phase transition around 150 K was found.\cite{Dai} It
seems that any new parent phase will initiate a series of new
superconductors by doping it away from the state with features of a
bad metal and the SDW order.

In this paper, we report the discovery of a new FeAs-based layered
compound SrFeAsF which has the ZrCuSiAs structure. As we know SrZnPF
is a compound with the ZrCuSiAs structure\cite{SrZnPF}. We replace
the ZnP sheets with FeAs sheets and get a new compound of SrFeAsF.
The compound SrFeAsF has the tetragonal space group P4/nmm at 300 K.
Both the resistivity and the dc magnetic susceptibility exhibit a
clear anomaly at about 173 K, which is attributed to the structural
and/or SDW transition. Surprisingly, a positive Hall coefficient
$R_\mathrm{H}$ has been found implying a dominant conduction by
hole-like charge carriers in this parent phase.

The SrFeAsF samples were prepared using a two-step solid state
reaction method, as used for preparing the LaFeAsO
samples.\cite{XYZ} In the first step, SrAs was prepared by reacting
Sr flakes (purity 99.9\%) and As grains (purity 99.99\%) at 500
$^o$C for 8 hours and then 700 $^o$C for 16 hours. They were sealed
in an evacuated quartz tube when reacting. Then the resultant
precursors were thoroughly grounded together with Fe powder (purity
99.95\%) and FeF$_3$ powder (purity 99\%) in stoichiometry as given
by the formula SrFeAsF. All the weighing and mixing procedures were
performed in a glove box with a protective argon atmosphere. Then
the mixture was pressed into pellets and sealed in a quartz tube
with an Ar atmosphere of 0.2 bar. The materials were heated up to
950 $^o$C with a rate of 120 $^o$C/hr and maintained for 60 hours.
Then a cooling procedure to room temperature was followed.

The dc magnetization measurements were done with a superconducting
quantum interference device (Quantum Design, SQUID, MPMS7). For the
magnetotransport measurements, the sample was shaped into a bar with
the length of 3 mm, width of 2 mm and thickness of about 0.9 mm. The
resistance and Hall effect data were collected using a six-probe
technique on the Quantum Design instrument physical property
measurement system (PPMS) with magnetic fields up to 9 T. The
electric contacts were made using silver paste with the contacting
resistance below 0.05 $\Omega$ at room temperature. The data
acquisition was done using a DC mode of the PPMS, which measures the
voltage under an alternative DC current and the sample resistivity
is obtained by averaging these signals at each temperature. In this
way the contacting thermal power is naturally removed. The
temperature stabilization was better than 0.1\% and the resolution
of the voltmeter was better than 10 nV.

\begin{figure}
\includegraphics[width=9cm]{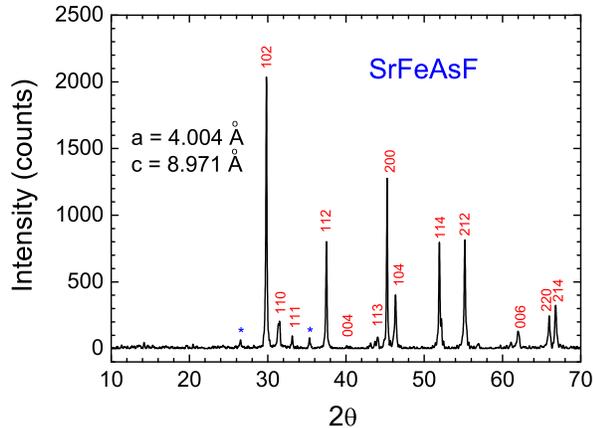}
\caption{(Color online) X-ray diffraction patterns for the SrFeAsF
sample. One can see that all the main peaks can be indexed to the
tetragonal ZrCuSiAs-type structure. The blue asterisks indicate the
little impurities from the SrF$_2$ phase. } \label{fig1}
\end{figure}

The X-ray diffraction (XRD) pattern for the sample SrFeAsF is shown
in Fig. 1. One can see that all the main peaks can be indexed to the
FeAs-1111 phase with the tetragonal ZrCuSiAs-type structure. Only
small amount of SrF$_2$ impurity phase was detected. By using the
software of Powder-X,\cite{DongC} we took a general fit to the XRD
data of this sample and the lattice constants were determined to be
a = 4.004 $\AA$ and c = 8.971 $\AA$. It is clear that the a-axis
lattice constant of this parent phase is slightly smaller than that
of the LaFeAsO system, while the c-axis one is much
larger,\cite{Kamihara2008,XYZ} indicating a completely new phase in
the present system since the radii of Sr$^{2+}$ is larger than that
of La$^{3+}$.

In Fig.2 (a) we present the temperature dependence of resistivity
for the SrFeAsF sample under magnetic fields up to 9 T. A rather
large value of the resistivity is observed. An upturn in the
low-temperature regime can be seen under all fields, representing a
weak semiconductor like behavior for the present sample. It is
unclear at this moment whether this behavior is intrinsic in nature,
or it is due to the weak localization effect, or some other effect.
This curve also reveals an anomaly at about $T_{an}$ = 173 K, which
may correspond to the structural and/or SDW transition, as has been
found in the parent phase of LnFeAsO (Ln = rare earth elements) and
(Ba, Sr)Fe$_2$As$_2$.\cite{Kamihara2008,BaKparent} Fig.2 (b) shows
the zero field cooled dc magnetization of the same sample at 5000
Oe. A clear anomaly at about 173 K in the magnetization curve
confirms the structural and/or SDW transition observed in the
resistivity data. Above 173 K, the magnetization exhibits a rough
linear temperature dependence, which may be a common effect in the
FeAs-based systems and was explained as due to short range
correlation of the local moments.\cite{ZhangGM}

\begin{figure}
\includegraphics[width=9cm]{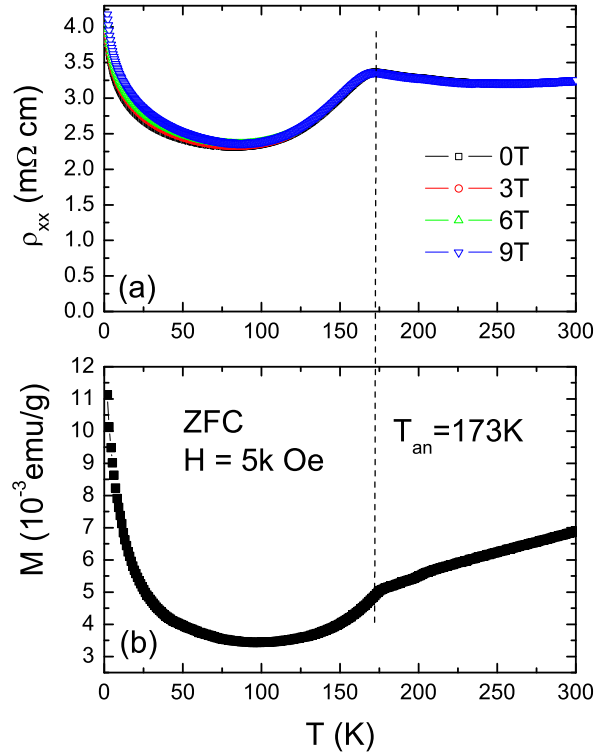}
\caption{(Color online) (a) Temperature dependence of resistivity
for the SrFeAsF sample under magnetic fields up to 9 T. A clear
anomaly at about $T_{an}$ = 173 K can be observed. (b) Temperature
dependence of dc magnetization for the zero field cooling (ZFC)
process at a magnetic field of $H$ = 5000 Oe. We can also see an
anomaly at the same temperature in the M(T) curve.} \label{fig2}
\end{figure}

\begin{figure}
\includegraphics[width=9cm]{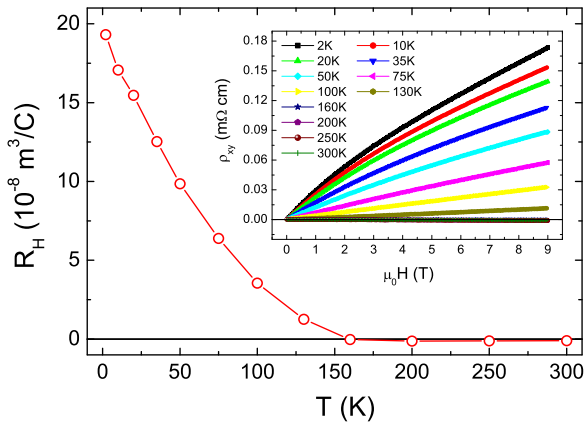}
\caption{(Color online)  Temperature dependence of Hall coefficient
$R_H$ determined on the present sample SrFeAsF. One can see a
monotonic decrease of $R_H$ in the temperature regime below about
160$\sim$ 170 K. Inset: The raw data of the Hall resistivity
$\rho_{xy}$ versus the magnetic field $\mu_0 H$ at different
temperatures. } \label{fig3}
\end{figure}

\begin{figure}
\includegraphics[width=8cm]{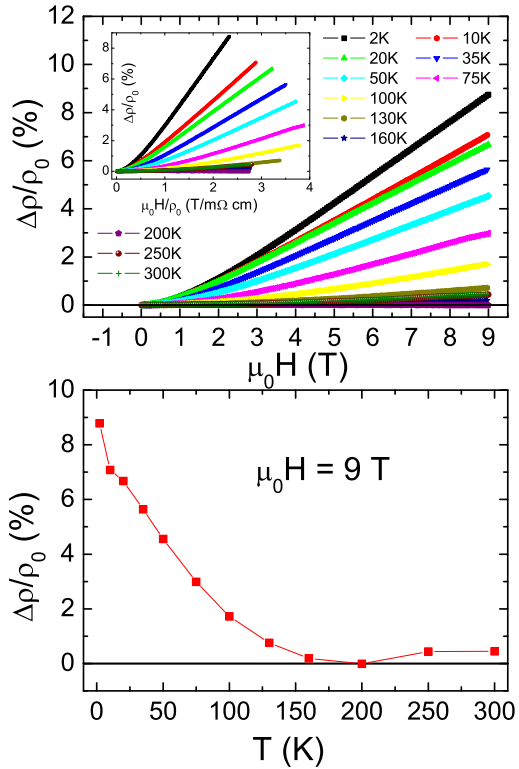}
\caption{(Color online) Field dependence of MR for the present
sample at different temperatures is shown in the top panel. A
moderate MR effect up to 9\% is observed under the field of 9 T at 2
K. Kohler plot of MR is presented in the inset. In the bottom panel
of this figure we show the temperature dependence of MR under the
field of 9 T. } \label{fig4}
\end{figure}

To get a comprehensive understanding to the conducting carriers in
the SrFeAsF phase, we measured the Hall effect of the present
sample. The inset of Fig. 3 shows the magnetic field dependence of
Hall resistivity ($\rho_{xy}$) at different temperatures. In the
experiment, $\rho_{xy}$ was taken as $\rho_{xy}$ = [$\rho$(+H) -
$\rho$(-H)]/2 at each point to eliminate the effect of the
misaligned Hall electrodes. A nonlinear field dependence of
$\rho_{xy}$ was observed in the temperature regime below 75 K, while
the linear behavior appeared above 100 K. This may suggest that a
multi-band effect or a complicated scattering mechanism (perhaps
magnetic related) emerged in the low temperature regime. The
temperature dependence of the Hall coefficient $R_H$ is presented in
the main frame of Fig.3. One can see that $R_H$ remains positive in
wide temperature regime and decreases monotonically in the
temperature regime below about 160$\sim$ 170 K and it becomes
slightly negative above that temperature. The sign changing of $R_H$
and the temperature dependent behavior may be related to the
structural and/or SDW transition as revealed by the resistivity
data, considering that this change occurred at temperatures close to
$T_{an}$. It is worth noting that the positive Hall coefficient
$R_H$ in this sample SrFeAsF is quite unique because in the two
parent phases of FeAs-based systems known so far, i.e., LnFeAsO (Ln
= rare earth elements) and (Ba, Sr)Fe$_2$As$_2$, the Hall
coefficient $R_H$ is negative. This strongly suggests that the
gapping to the Fermi surfaces induced by the SDW order is more
complex than we believed before, and the case in SrFeAsF is that it
removes the density of states on some Fermi pockets and may leave
one of the hole pockets partially or fully ungapped. Our data
clearly show that it is possible for the parent phase to have
electron-like or hole-like charge carriers. It is well known that in
the conventional metals the Hall coefficient $R_H$ is almost
independent of temperature. The strong temperature dependence of
$R_H$ below $T_{an}$ in our data suggests either a strong multi-band
effect or the variation of the charge carrier densities, or both
effects collectively contribute to the Hall signal in the present
parent phase of SrFeAsF.

The magnetoresistance (MR) is a very powerful tool to investigate
the properties of electronic scattering.\cite{LiQ,YangHall} Field
dependence of MR for the present sample at different temperatures is
shown in the main frame in the top part of Fig 4. One can see a
moderate MR effect up to 9\% under the field of 9 T at 2 K. This is
a rather large magnitude compered with the F-doped LnFeAsO
samples.\cite{XYZ,NdSingle} The semiclassical transport theory has
predicted that the Kohler's rule will be held if only one isotropic
relaxation time is present in a solid state system.\cite{Kohler} The
Kohler's rule can be written as
\begin{equation}
\frac{\Delta \rho}{\rho_0}=\frac{\rho(H)-\rho_0}{\rho_0}=F(\frac{
H}{\rho_0}),\label{eq:1}
\end{equation}
where $\rho(H)$ and $\rho_0$ represents the longitudinal resistivity
at a magnetic field $H$ and that at zero field, respectively.
Equation (1) means that the $\Delta \rho/\rho_0$ vs $ H/\rho_0$
curves for different temperatures, the so-called Kohler plot, should
be scaled to a universal curve if the Kohler's rule is obeyed. The
scaling based on the Kohler plot of our sample is revealed in the
inset of the top part of Fig.4. An obvious violation of the Kohler's
rule can be seen from this plot. This behavior may indicate a
multi-band effect or a gradual gapping effect to the density of
states by the SDW ordering in the present sample. Temperature
dependence of MR under the field of 9 T is shown in the bottom part
of Fig 4. Rather similar to that observed in the $R_H$ vs $T$ plot,
$\Delta \rho/\rho_0$ decreases monotonically in the low temperature
regime below about 200 K and a minimum appears around  $T_{an}$.
This may provide another evidence of the influence of the structural
and/or SDW transition on the behavior of the conducting charge
carriers.

In summary, a parent phase, namely SrFeAsF,  with the ZrCuSiAs
structure was synthesized successfully using a two-step solid state
reaction method. An anomaly at about 173 K can be observed from the
data of the resistivity and dc magnetization, which is ascribed to
the structural and/or SDW transition. Also strong Hall effect and
moderate MR were observed below $T_{an}$. We found that the Hall
coefficient $R_H$ is positive below $T_{an}$, displaying an opposite
behavior comparing to the cases in the two parent phases of
FeAs-based systems known so far, i.e., LnFeAsO (Ln = rare earth
elements) and (Ba, Sr)Fe$_2$As$_2$ where the Hall coefficient $R_H$
is negative. This suggests that the gapping to the Fermi surfaces
induced by the SDW order may remove the density of states on some
Fermi pockets and leave one of the hole pockets partially or fully
ungapped in the present parent phase. Our results clearly show that
it is possible for the parent phase to have electron-like or
hole-like charge carriers. We also observed a moderate
magnetoresistance up to 9\% under the field of 9 T. The violation of
the Kohler's rule along with the strong temperature dependence of
$R_H$ may suggest a multi-band and/or a spin scattering effect in
this system. By doping strontium with lanthanum, we found
superconductivity in Sr$_{1-x}$La$_x$FeAsF, which will be presented
separately.\cite{XYZ2}

Note added: When we were finalizing this paper, we became aware that
a paper was posted on the website on the same day of our submission.
That paper reports also the synthesizing of the compound SrFeAsF and
a different set of data.\cite{Johrendt}

This work is supported by the Natural Science Foundation of China,
the Ministry of Science and Technology of China (973 project:
2006CB01000, 2006CB921802), the Knowledge Innovation Project of
Chinese Academy of Sciences (ITSNEM).

\end{document}